# To the methodology of low temperature technologies: mathematical analysis of Langevin forces in superconductor

Iogann Tolbatov[*]

*Physics and Engineering Department, Kuban State University, Krasnodar, Russia*

Langevin forces are investigated in the framework of the phenomenological Ginzburg-Landau (GL) theory. These stochastic forces introducted in the time-dependent Ginzburg-Landau (TDGL) equation describing the superconductor transport properties above the critical temperature model the fluctuations action. We assume that there exists a profound connection between the fluctuation Cooper pair energy spectrum and the Langevin forces possible values spectrum. In investigation carried out on the basis of that hypothesis we obtain the analytical expression for Langevin forces defining them by means of order parameter, i.e., by means of the fluctuation Cooper pair wave function. Langevin forces properties are analyzed. The conlusion about their complex nature is done. Asymptotic analysis of Langevin forces is performed.

PACS number: 74.40.+k

## I. INTRODUCTION

With the help of the stochastic differential equation later named after him, Langevin[1] described in 1908 the Brownian motion, i.e., the persistent chaotic motion of small particles suspended in liquid or gas. This motion type peculiarity consists in its unlimitedly long duration without any visible alteration. The Brownian particles motion intensity depends on their dimensions, but it does not depend on their nature. This motion intensity increases with the temperature increase and with the liquid or gas viscosity decrease. The Brownian motion is not molecular motion. It is an immediate proof of the molecules existance and their thermal motion chaotic character. The Brownian motion is caused by the fluctuations of pressure on the small particle surface from the solvent molecules.

## II. LANGEVIN FORCE

Equation studied by Langevin describes the Brownian motion with a constant potential. Acceleration $\vec{a}$ of the Brownian particle of a mass $m$ is defined by means of the sum of the viscous friction force, which is proportional to the particle velocity $\vec{v}$ (the Stoke's law), and the noise term $\vec{\eta}(t)$ (the name, which is used in physics for to mean the stochastic process in differential equation), as the result of the persistent collisions of the particle with the liquid molecules, and the systematic force $\vec{\Phi}(x)$ appearing due to the inside- and outside- molecular interactions:

$$m\vec{a} = m\frac{d\vec{v}}{dt} = \vec{\Phi}(x) - \gamma\vec{v} + \vec{\eta}(t). \tag{1}$$

If the external forces are not taken into account, and if we consider only one coordinate, since it does not lead to the generality loss, Eq. (1) transforms into the form:

$$m\ddot{x} = -\frac{1}{B}\dot{x} + F(t). \tag{2}$$

We also consider that the random force $F(t)$ answers the following conditions:

$$\langle F(t)\rangle = 0, \tag{3}$$

$$\langle F_i(t_1)F_j(t_2)\rangle = b\delta_{ij}\delta(t_1 - t_2), \tag{4}$$

where $b$ is some constant value, $\delta_{ij}$ is the Kronecker symbol, $\delta(t_1 - t_2)$ is the Dirac delta function. Angle brackets mean the average over time. The resultant force $\vec{F}$ of molecules strikes on particle from different sides has a random character, and it is called the Langevin force.

Langevin force properties were not practically being studied because of the stochastacity of differential equation, in which it was introduced. Since Langevin forces answer the fluctuation-dissipation theorem,[2] they are to be correlated by means of the Gaussian white-noise law. It can be written down in the form:

$$\langle F(t_1)F(t_2)\rangle = q\delta(t_1 - t_2), \tag{5}$$

where the Langevin force spectral density $S(\omega)$, which by the Wiener-Khintchine theorem is the Fourier transform of the correlation function Eq. (5), is independent of the frequency $\omega$:

$$S(\omega) = 2\int_{-\infty}^{\infty} e^{-i\omega t} q\delta(\tau)d\tau = 2q. \tag{6}$$

Exactly due to this frequency $\omega$ independency, the Langevin force Eq. (5) with a $\delta$ correlation is called white-noise force. Really, how can one investigate the force, the nature of which is chaotic, the force, which is introduced in the equation for to model the chaos action? Nevertheless, we consider that it is interesting and necessary to analyze the Langevin forces action on the fluctuation Cooper pair for to define its properties and its energy possible values spectrum, since the fluctuation Cooper pair wave function is the connecting parameter in the phenomenological GL theory.

### III. LANGEVIN FORCES AND LANGEVIN EQUATION IN THE THEORY OF FLUCTUATIONS IN SUPERCONDUCTORS

The fluctuation Cooper pairs appearance in superconductor at temperatures above the superconducting transition temperature leads to the opening of charge transfer new channel. Their singular contribution to the conductivity in the critical temperature vicinity is called the paraconductivity.[3]

For to find the paraconductivity the time is included in the GL scheme, since the conductivity determines the system response to the applied electric field defined as the vector-potential time-derivative: $E = -\frac{\partial A}{\partial t}$, where the vector-potential $A$ and the fluctuating order parameter are offered to be considered as dependent not only on coordinates, but also on time.[4]

The stationary GL theory equations can be applied under condition of the small deviations from the equilibrium and if during the order parameter relaxation the time-derivative of the latter $\frac{\partial \Psi}{\partial t}$ will be proportional to the variation derivative of the free energy GL functional $\frac{\delta F}{\delta \Psi^*}$. The model TDGL equation is written down in the form:

$$-\gamma_{GL}\left(\frac{\partial}{\partial t} + 2ie\varphi\right)\Psi = \frac{\delta F}{\delta \Psi^*} + \zeta(r,t), \qquad (7)$$

where $\varphi$ is the scalar potential of the electric field, $\gamma_{GL}$ is the dimensionless complex coefficient defined by means of the fluctuation Cooper pair life time $\tau_{GL}$,

$$\operatorname{Re}\gamma_{GL} = \alpha T_C \varepsilon \tau_{GL} = \frac{\pi\alpha}{8}, \quad \gamma_{GL} = \frac{\pi\alpha}{8} + i\operatorname{Im}\gamma_{GL}.$$

One can take into account the thermodynamical fluctuations with the help of the Langevin forces $\zeta(r,t)$, which model the fluctuations influence on superconductor, being introduced in the right-hand-side of Eq. (7), which characterizes the order parameter relaxation dynamics.

In the TDGL equation the Langevin forces also are to answer the fluctuation-dissipation theorem.[2] The problem of definition of Langevin forces acting on fluctuation Cooper pair possible values spectrum was not set, though the Langevin equation and, in particular, its numerical simulations for the solution of various problems of the superconductivity physics were used.

In investigation carried out by Aranson, Koshelev, and Vinokur[5] the disorder-induced effective Langevin force was determined by means of establishing of the universal structure of the phase-transition lines as functions of renormalized interlayer coupling and "shaking temperature". But it was done for to determine the phase diagram for a driven vortex lattice in layered superconductors. A successful investigation was carried out through the numerical simulations of the time-dependent Ginzburg-Landau-Lawrence-Doniach equations.

The Langevin equation was used for to construct a self-consistent theory describing the vortex dynamics in disordered superconductors by Grundberg and Rammer.[6] The validity of this theory was ascertained by comparing with numerical simulations of the Langevin equation.

Langevin simulations were used for to examine the colloidal dynamics on disordered substrates in the perfect investigation of Reichhardt and Olson.[7] There Langevin simulations were used for to examine driven colloids interacting with quenched disorder.

Langevin's approach of the stochastic differential equation was used by Mishonov *et al.*[8] for the derivation of the Boltzman kinetic equation for fluctuation Cooper pairs.

The two-dimensional Langevin simulations were performed for to study the peak effect of the critical current of type II superconductors with strong magnetic pinning by Xu *et al.*[9] The simulations revealed there that the peak-effect could take place for certain pinning strengths, densities of pinning centers, and driving forces only. And this fact agreed precisely with experiments.

## IV. INVESTIGATION HYPOTHESIS

We can not observe any enthusiasm in studying Langevin forces on the part of investigators. In our opinion, there is a sense in studying them, notwithstanding their stochasticity, because just they express the thermodynamical fluctuations effect on the order parameter. And since exactly the order parameter represents the fluctuation Cooper pair wave function in the GL methodology framework, we assume that it is possible to find the Langevin forces acting on one fluctuation Cooper pair. And when the Langevin forces expression will be obtained, then studying these stochastic forces properties, we investigate at the same time the properties of the fluctuation Cooper pair.

Our investigation hypothesis assumes the existance of profound connection between the properties of the fluctuation Cooper pair and these of the Langevin forces acting on it. According to this hypothesis, it is the coupled electrons energy spectrum that plays the most significant role in definition of the Langevin forces general form in the TDGL equation. Existance of imaginary part of these forces also, in our opinion, is an interesting and important problem significant in a practical sence for the fluctuation Cooper pair motion modeling. Therefore, the scientific novelty consists not only in the problem solution method, but also in the problem setting.

Our investigation is carried out for the case, when the superconductor is in the vicinity of the superconducting transition and its temperature is close to the critical: $\varepsilon \ll 1$ ($\varepsilon$ is the reduced temperature), i.e., the considered conditions do not contradict to the GL theory applicability conditions.

## V. INVESTIGATION OF LANGEVIN FORCES ACTING ON FLUCTUATION COOPER PAIR

At the first stage, we study the Langevin forces effect on the fluctuation Cooper pair in the electric field absence. We suggest to define them by means of the order parameter $\Psi^{(0)}(r,t)$.

Neglecting the fourth order term in the GL functional, we write down Eq. (7) in operator form:[10]
$$\left[\hat{L}^{-1} - 2ie\gamma_{GL}\varphi(r,t)\right]\Psi(r,t) = \zeta(r,t), \tag{8}$$
where the operator $\hat{L}$ and Hamiltonian $\hat{H}$ are determined as:

$$\hat{L} = \left[\gamma_{GL}\frac{\partial}{\partial t} + \hat{H}\right]^{-1}, \quad \hat{H} = \alpha T_C\left[\varepsilon - \hat{\xi}^2(\hat{\nabla} - 2ieA)^2\right], \quad (9)$$

where $\varepsilon = \ln\frac{T}{T_C}$ is the reduced temperature.

In the electric field absence one can find the formal solution of Eq. (8) as:
$$\Psi^{(0)}(r,t) = \hat{L}\zeta(r,t). \quad (10)$$

The Langevin forces must satisfy the fluctuation-dissipation theorem, which means that the correlator $\langle \Psi_p^{(0)*}(t')\Psi_p^{(0)}(t)\rangle$ at the coinciding moments of time has to transform into the value $\langle |\Psi_p|^2\rangle$ found by means of the averaging of the order parameter modulus square over fluctuations in thermal equilibrium:

$$\langle |\Psi_p|^2\rangle = \frac{\int D\Psi_p D\Psi_p^* |\Psi_p|^2 \exp\{-\alpha(\varepsilon + \xi^2 p^2)|\Psi_p|^2\}}{\int D\Psi_p D\Psi_p^* \exp\{-\alpha(\varepsilon + \xi^2 p^2)|\Psi_p|^2\}} = \frac{1}{\alpha(\varepsilon + \xi^2 p^2)}. \quad (11)$$

This requirement is fulfilled in the case if the Langevin forces $\zeta(r,t)$ and $\zeta^*(r,t)$ are correlated by the Gaussian white-noise law:[10]

$$\langle \zeta^*(r,t)\zeta(r',t')\rangle = 2T\,\mathrm{Re}\,\gamma_{GL}\delta(r-r')\delta(t-t'). \quad (12)$$

For to show this, we must limit ourselves to the case, when the vector-potential $A$ equals zero, so we calculate the order parameter correlator $\langle \Psi_p^{(0)*}(t')\Psi_p^{(0)}(t)\rangle$:

$$\langle \Psi^*(r,t)\Psi(r',t')\rangle = \langle \zeta^*(r,t)\hat{L}^*\hat{L}\zeta(r',t')\rangle = 2T\,\mathrm{Re}\,\gamma_{GL}\int\frac{dp}{(2\pi)^D}e^{ip(r-r')}\int_{-\infty}^{\infty}\frac{d\Omega}{2\pi}L^*(p,\Omega)L(p,\Omega). \quad (13)$$

The fundamental solution of Eq. (13) $L(p,\Omega)$ can be found through the Fourier transformation of Eq. (9), which leads the latter to the form:
$$L(p,\Omega) = (-i\gamma_{GL}\Omega + E_p)^{-1}. \quad (14)$$

The expression Eq. (14) substitution in the order parameter correlator Eq. (13) and the integral over frequencies calculation finish the proof of the fact that the order parameter correlator $\langle \Psi_p^{(0)*}(t')\Psi_p^{(0)}(t)\rangle$ in the coinciding moments of time transforms into $\langle |\Psi_p|^2\rangle$:

$$\langle \Psi^*(r,t)\Psi(r',t)\rangle_p = 2T\,\mathrm{Re}\,\gamma_{GL}\int_{-\infty}^{\infty}\frac{d\Omega}{2\pi}\frac{1}{|-i\gamma_{GL}\Omega + E_P|^2} = \langle |\Psi_p|^2\rangle, \quad (15)$$

where
$$E_p = \alpha T_C(\varepsilon + \hat{\xi}^2 p^2) \quad (16)$$
is the fluctuation Cooper pair energy spectrum.

At the second stage of our investigation, we consider the formal correlation length operator $\hat{\xi}$, which is used in Eq. (16):

$$\hat{\xi} = -\frac{\nabla^2}{2m} + U_i(r), \quad (17)$$

$$\hat{\xi}^2 = \left(\frac{\nabla^2}{2m}\right)^2 - 2\left(\frac{\nabla^2}{2m}\right)U_i(r) + U_i^{\,2}(r).$$

$$\hat{\xi}^2 p^2 = \left(\frac{\nabla^2}{2m}\right)^2 p^2 - 2\left(\frac{\nabla^2}{2m}\right)pU_i(r)p + U_i^2(r)p^2. \tag{18}$$

In the magnetic field absence one can rewrite the expression Eq. (18) in the form:

$$\hat{\xi}^2 p^2 = \left(\frac{\nabla^2}{2m}\right)^2 p^2 = E^2(p), \tag{19}$$

where $E(p)$ is the fluctuation Cooper pair energy.

At the third stage of investigation, we find formally the Langevin forces from the solution Eq. (9) of the TDGL equation Eq. (8):

$$\zeta(r,t) = \hat{L}^{-1}\Psi^0(r,t). \tag{20}$$

Since we consider the Langevin forces influencing on the fluctuation Cooper pair, taking into account Eq. (14), we can rewrite Eq. (20) in the form:

$$\zeta(r,t) = \Psi^0(r,t)(E_p - i\gamma_{GL}\Omega), \tag{21}$$

where $\Omega$ is the frequency characteristics.

Using the fluctuation Cooper pair energy spectrum Eq. (16), we obtain the expression for the Langevin forces influencing on the fluctuation Cooper pair:

$$\zeta(r,t) = \Psi^0(r,t)(\alpha T_C(\varepsilon + \hat{\xi}^2 p^2) - i\gamma_{GL}\Omega). \tag{22}$$

Since we act in the GL theory applicability region ($\varepsilon \ll 1$), we can write down that:

$$\varepsilon = \ln\frac{T}{T_C} \approx \frac{T - T_C}{T_C}.$$

Consequently, the expression Eq. (22) transforms into the form:

$$\zeta(r,t) = \Psi^0(r,t)\left(\alpha T_C\left(\frac{T - T_C}{T_C}\right) + \alpha T_C E^2(p) - i\gamma_{GL}\Omega\right). \tag{23}$$

Hence, we find the expression for the Langevin forces influencing on the fluctuation Cooper pair in the electric field absence:

$$\zeta(r,t) = \Psi^0(r,t)(\alpha T + \alpha T_C(E^2(p) - 1) - i\gamma_{GL}\Omega). \tag{24}$$

## VI. ASYMPTOTIC ANALYSIS OF EXPRESSION FOR LANGEVIN FORCES

At the fourth stage, we investigate the expression Eq. (24). It follows from it that the Langevin forces are complex. Let us consider separately their real

$$\text{Re}\,\zeta(r,t) = \Psi^0(r,t)(\alpha T + \alpha T_C(E^2(p) - 1)) \tag{25}$$

and imaginary

$$\text{Im}\,\zeta(r,t) = \Psi^0(r,t)(-\gamma_{GL}\Omega) \tag{26}$$

parts.

The imaginary part is in its minimum (tends to zero) if the frequency characteristics of the Langevin forces influencing on the fluctuation Cooper pair in the electric field absence tends to zero: $\Omega \to 0$.

The real part is minimum if it is equal to zero. The coefficient $\alpha$ can not be equal to zero $\alpha \neq 0$. The order parameter $\Psi^0(r,t)$, i.e., the fluctuation Cooper pair wave function, also equals zero.

Hence, the Langevin forces real part $\text{Re}\,\zeta(r,t)$ equals zero if:

$$T + T_C(E^2(p) - 1) = 0, \tag{27}$$

i.e., if $E^2(p) = -\dfrac{T - T_C}{T_C} = -\ln\dfrac{T}{T_C} = -\varepsilon$.

With the fluctuation Cooper pair energy $E(p) = i\sqrt{\varepsilon}$, by the way, imaginary in the electric field absence and with frequencies $\Omega \to 0$, the Langevin forces are minimum.

The Langevin forces are maximum by the fluctuation Cooper pair energy $E(p) \to \infty$.

## VII. OUR INVESTIGATION RESULTS AND THEIR INTERPRETATION

Thus, at the four stages of investigation we have expressed the Langevin forces acting on the fluctuation Cooper pair in the electric field absence by means of the Cooper pair wave function (the so-called order parameter) using its energy spectrum defined by the TDGL equation calculation. This permits us to describe the Langevin forces in the following method:
  (1) The Langevin forces are complex, they consist of both the real Eq. (25) and imaginary Eq. (26) parts.
  (2) The Langevin forces are maximum if the fluctuation Cooper pair energy is $E(p) \to \infty$. They are minimum by the frequency characteristics $\Omega \to 0$ and by the fluctuation Cooper pair energy $E(p) \to i\sqrt{\varepsilon}$. We must notice that the energy, which is necessary to the Langevin forces minimum values, is imaginary.

Our investigation gives a full characteristics of the Langevin forces modeling the fluctuations action in the TDGL equation. In the study we have not violated the GL theory applicability frames, and this is very important, it proves the carried out investigation correctness. The scientific novelty consists both in the problem formulating and in its solution. Theoretical significance of the conducted study is especially distinctly seen while highly untrivial obtained results interpreting. The discovered evident complex nature of the Langevin forces mathematical expression does not have any physical sence except of the case of its multiplication by its complex conjugate – when its imaginary part disappears. This peculiarity is noticeable in the Gaussian white-noise law Eq. (12), where the averaged product of the conjugated stochastic forces is used.

The defined asymptotics of expression for Langevin forces throws light upon some of their properties. While the statement about their maximum values, as the fluctuation Cooper pair energy, on which they act, becomes infinite, is somewhat trivial, these forces tending to minimum values at the finite nonzero value of the fluctuation Cooper pair turns out to be absolutely unexpectable. The nonstationary nature of the fluctuation transport is the cause of "the potential well" existance. The fluctuation Cooper pair behavior in this potential field can be interpreted as the Heisenberg uncertainty principle result: the fall on the potential well bottom is connected with the fluctuation Cooper pair momentum transformation into zero, and at the same time the momentum uncertainty turns into zero. Hence, the fluctuation Cooper pair coordinate uncertainty becomes infinitely great, what contradicts in its turn to the particle stay in "the potential well". Exactly that is why, the fluctuation Cooper pair minimum energy does not equal zero. The existance of the Cooper pair minimum nonzero energy can be also interpreted as the fluctuation Cooper pair energy uncertainty in the nonstationary fluctuation transport process.

## VIII. CONCLUSION

We have investigated the Langevin forces, which were introduced by Ginzburg and Landau in the time-dependent equation for to model the fluctuation action. Their properties analysis justified itself and gave us new significant results, such as the formula connecting the Langevin forces with the order parameter Eq. (24), and the complex nature of these stochastic forces Eq. (25) and Eq. (26). Our investigation was carried out on the phenomenological level. Nevertheless, we consider as a perspective the study of these forces in the framework of the microscopic theory of the nonstationary Bardeen-Cooper-Schrieffer equations for the discovery of the fluctuation transport new peculiarities. Because exactly these forces influence is the most significant on the charge transfer channel characteristics at temperatures above the critical temperature.